\documentclass[twocolumn,showpacs,showkeys,preprintnumbers,prl,amsmath,amssymb]{revtex4}

\usepackage[dvips]{graphicx}
\usepackage{dcolumn}
\usepackage{bm}

\newcommand{\Prb}{{\em Phys. Rev. B} }
\newcommand{\Prl}{{\em Phys. Rev. Lett.} }
\newcommand{\Apl}{{\em Appl. Phys. Lett.} }
\newcommand{\rstd}{$R_{std}$ }
\newcommand{\rstdy}{$R_{std}$}

\newcommand{\bite}{Bi$_2$Te$_3$/FeTe }

\newcommand{\bt}{Bi$_2$Te$_3$ }

\newcommand{\be}{\begin{equation} }

\newcommand{\bib}{\bibitem}

\newcommand{\cmcy}{cm$^{3}$}
\newcommand{\cms}{cm$^{2}$ }

\newcommand{\ene}{\end{equation}}

\newcommand{\figr}{Figure~\ref}

\newcommand{\hc}{$B_{c}$ }
\newcommand{\hcl}{$B_{c1}$ }

\newcommand{\hcu}{$B_{c2}$ }

\newcommand{\hcuy}{$B_{c2}$}
\newcommand{\hcy}{$B_{c}$}

\newcommand{\jj}{$j$ }
\newcommand{\jjy}{$j$}
\newcommand{\jc}{$j_{c}$ }

\newcommand{\jd}{$j_{d}$ }
\newcommand{\jdt}{$j_{d}(T)$ }

\newcommand{\jdy}{$j_{d}$}

\newcommand{\lamy}{$\lambda_L$}
\newcommand{\lm}{$\lambda$ }
\newcommand{\lmy}{$\lambda$}

\newcommand{\mgb}{MgB$_{2}$ }
\newcommand{\mgby}{MgB$_{2}$}

\newcommand{\nsy}{$n_{s}$}

\newcommand{\psay}{$|\psi|$}

\newcommand{\rrho}{{$\rho$} }
\newcommand{\rrhon}{$\rho_{n}$ }
\newcommand{\rrhony}{$\rho_{n}$}
\newcommand{\rrhoy}{{$\rho$}}
\newcommand{\rrhos}{{$\rho_s$} }
\newcommand{\rrhosy}{{$\rho_s$}}

\newcommand{\scv}{superconductivity }
\newcommand{\scvy}{superconductivity}
\newcommand{\sg}{superconducting }

\newcommand{\scg}{superconducting }

\newcommand{\ssc}{superconductor }
\newcommand{\sscy}{superconductor}

\newcommand{\tc}{$T_{c}$ }

\newcommand{\tcy}{$T_{c}$}

\newcommand{\ybco}{Y$_{1}$Ba$_{2}$Cu$_{3}$O$_{7}$ }
\newcommand{\ybcoy}{Y$_{1}$Ba$_{2}$Cu$_{3}$O$_{7}$}
\newcommand{\ncco}{Nd$_{2-x}$Ce$_{x}$CuO$_{4}$ }

\newcommand{\row}{\rightarrow }

\begin{document}
\preprint{Condens. Matter {\bf 4}, 54 (2019). Review article in 
Special Issue "From Cuprates to Room Temperature Superconductors"}
\title{Evaluating Superconductors through Current 
Induced Depairing}

\author{Milind N. Kunchur} 
\email[Corresponding author email: ]{kunchur@sc.edu} 
\homepage{http://www.physics.sc.edu/~kunchur}


\affiliation{Department of Physics and Astronomy, University of South
Carolina, Columbia, SC 29208}


\begin{abstract} The phenomenon of \scv occurs in the phase space of three principal parameters: temperature T, magnetic field B, and current density \jjy . The critical temperature \tc is one of the first parameters that is measured and in a certain way defines the \sscy. From the practical applications point of view, of equal importance is the upper critical magnetic field \hcu and conventional critical current density \jc (above which the system begins to show resistance without entering the normal state). However, a seldom-measured parameter, the depairing current density \jdy , holds the same fundamental importance as \tc and \hcuy , in that it defines a boundary between the \scg and normal states. A study of \jd sheds unique light on other important characteristics of the superconducting state such as the superfluid density and the nature of the normal state below \tcy , information that can play a key role in better understanding newly-discovered \sg materials. 
From a measurement perspective, the extremely high values of \jd make it difficult to measure, which is the reason why it is seldom measured. Here, we will review the fundamentals of current-induced depairing and the fast-pulsed current technique that facilitates its measurement and discuss the results of its application to the topological-insulator/chalcogenide interfacial \sg system.
\end{abstract} 



\maketitle
The phenomenon of \scv has a long and rich history: from the initial discovery in 1911 of \scv in mercury at liquid-helium temperature \cite{onnes} to the recent discovery of room-temperature \scv in lanthanum superhydride \cite{soma,drozdov}. Numerous parameters, probed by a variety of techniques, are used to characterize the \scg state. However, the mixed-state upper critical field \hcuy , reflective of the coherence length $\xi$, and the penetration depth \lamy, reflective of the superfluid density \rrhosy =$1/\lambda^2$, are two crucial measurements that are amongst the first to be performed. There are multiple techniques for determining such parameters, each of which has its own advantages and limitations. Our group has developed some uncommon, and in some cases unique, experimental techniques that investigate \sscy s at ultra-short time scales, and under unprecedented and extreme conditions of current density \jjy , electric fields $E$, and 
power density $p=\rho j^2$ (where \rrho is the resistivity). These techniques have led to the discovery or confirmation of several novel phenomena and regimes in \sscy s and in addition provide an alternative method to glean information on fundamental \scg parameters, which in some cases may be hard to obtain by other methods. 
These methods and approaches are highly relevant in the search for new \scg materials and in developing an understanding of their fundamental properties. This article discusses the physical meaning 
of \jd and its interrelationships with other basic parameters of the \sg state, as 
well as the technical challenges in measuring this important critical parameter. We discuss our results from this approach in the study of the topological-insulator/chalcogenide interfacial \sg system.

\section{Introduction} 
Attractive interactions 
between charge carriers cause them to condense by pairs into a coherent macroscopic quantum state below some transition temperature \tcy . 
The formation of this state 
is governed principally by a competition between four energies: condensation,
magnetic-field expulsion, thermal, and kinetic. The order
parameter $\Delta$, which
describes the extent of condensation and the strength of the 
superconducting state, is
reduced as the temperature $T$, magnetic field $B$, and electric current
density $j$ are increased. In type-II \sscy s, there is partial flux entry at $B$ values above the lower critical magnetic field \hcl and complete destruction of \scv above the upper critical field \hcu (type-I \sscy s can be viewed as a special case where the thermodynamic critical field \hc= \hcl= \hcuy). 
The boundary in the $T$-$B$-$j$ phase space that separates the 
superconducting and normal states
is where $\Delta$ vanishes, and the three parameters attain their
critical values $T_{c2}(B,j)$,
$B_{c2}(T,j)$, and $j_{d}(T,B)$. 
$j_{d}$ sets the intrinsic upper limiting scale for 
supercurrent transport in any superconductor, and for $j > j_d$, the system attains its normal-state resistivity $\rho_n$. \jd should be distinguished from the conventional critical value \jc (related to extrinsic characteristics such as the depinning of vortices) above which there is partial resistivity $\rho < \rho_n$.

The resistivity \rrho in the \scg state is usually less than its normal-state value \rrhony. The reason for the presence of resistance at all in the \scg state is because of fluctuations, percolation through junctions (in the case of granular \sscy s), and the motion of magnetic flux vortices. 
For singly-connected \sscy s not very close to \tcy , only the last mechanism dominates as the cause of resistance. 
In the magnetic field region between \hcl and \hcuy , a type II \ssc enters a ``mixed state'' with quantized magnetic flux vortices, each containing an elementary quantum of flux $\Phi_0=h/2e$. Under the Lorentz driving force of an applied current, $\bm j$$\times$$\Phi_0$, vortices move transverse to $j$, leading to a flux-flow resistivity: 
\be \label{fff} 
\rho_f \sim \rho_n B/B_{c2}
\ene
in the free-flux-flow (large driving force) limit. 

Two length scales characterize the \scg state \cite{tinkhamtext}. One is the coherence length: 
\be 
\xi = v_F \tau_{\Delta} \simeq \hbar v_F/\pi\Delta 
\ene
which is the characteristic length scale for spatial modulations in $\Delta$ 
(here, $v_F$ is the Fermi velocity and $\tau_{\Delta}$ is the order-parameter relaxation time). The normal core of a flux vortex has an approximate effective radius of $\xi$. The destruction of the \scg state occurs when these normal cores overlap, corresponding to the condition:
\be \label{bc2-xi} 
B_{c2} = \frac{\Phi_0}{2\pi \xi^2} = \frac{\Phi_0}{2\pi \xi_1 \xi_2}
\ene
where $\xi$ is the coherence length perpendicular to $B$; the single $\xi$ is replaced by the geometric mean 
$\sqrt{\xi_1 \xi_2}$ in cases where the plane perpendicular to $B$ is characterized by two anisotropic values. 

\begin{figure}[ht]
\includegraphics[width=0.85\hsize]{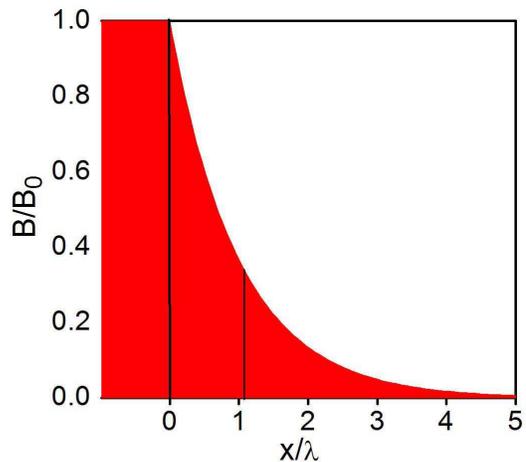} 
\caption{An externally-applied magnetic field $B_0$ is screened from the interior of a superconductor in the Meissner state over a characteristic length scale, which is the magnetic-field penetration depth $\lambda$: $B \sim B_0 e^{-x/\lambda}$. The circulating screening current is of roughly the depairing magnitude \jdy, so that $\lambda \propto 1/j_d$.}
\label{lambda-screening} 
\end{figure}

The other characteristic length scale in a superconductor 
is the magnetic-field penetration depth $\lambda$, whose London value is given by: 
\be \label{Ll}
\lambda_L = \sqrt{\frac{m^*}{\mu_0 n_s e^2}} 
\ene
where $m^*$ is the effective electronic mass and $n_s$ is the density of superconducting electrons. The theory behind this important quantity and its relationship to \jd is described below. \figr{lambda-screening} shows the profile of the magnetic field as it gets screened from the interior of a superconductor. 

\subsection{Superfluid Density} 
For clean metallic superconductors, $n_s \row n$ as $T \row 0$, where $n$ is the concentration of carriers in the normal state. 
Non-local effects and other corrections lead to deviations in \lm from its London value. 
Hence, the superfluid density $\rho_s$ can be conveniently and more completely defined as: 
\be \rho_s \equiv 1/\lambda^2
\ene
which includes the effective mass and other corrections to the effective $n_s$, 
rather than the simpler definitions $\rho_s = n_s$ or $\rho_s = n_s/m^*$ that are sometimes used in the literature.
\rrhosy , a quantity of central importance in \scvy , 
characterizes the phase stiffness of the condensate \cite{emery_kivelson_nature} and its 
effectiveness at screening out magnetic fields and feeds into expressions for the transition temperature (such as the 
Uemura relation $T_c \propto \rho_s(0)$ that applies to the underdoped \linebreak cuprate \sscy s \cite{Uemura1,Uemura2,Hardy}). 

Traditionally, many common methods for obtaining \rrhos do so by directly or indirectly measuring \lm through its effect on a superconducting sample's magnetic-field profile and consequent magnetic susceptibility. This category includes methods such as reflection of spin-polarized slow \linebreak neutrons \cite{Felcher}, mutual inductance altered by an intervening \scg film \cite{Claassen,yong-lemberger}, 
 changes in the self-inductance of a coil that is part of an LC resonating circuit \cite{Boghosian,Degrift}, muon-spin rotation \cite{Sonier}, magnetic force microscopy \cite{Luan}, microwave cavity resonance \cite{Shibauchi}, and measurements of the lower critical field. 
These measurements are understandably affected if the material's internal magnetic field is altered, for example by a large paramagnetic background as in the case of the \ncco \ssc because of its Nd$^{3+}$ magnetic moments. 

Another approach to obtaining \rrhos is by measuring the inertia of the superfluid (kinetic inductance) during its ballistic acceleration phase \cite{lee-lemberger,ballistic,Diener,Jochem}. This method requires the sample to be patterned into very high aspect ratio 
meanders for the highest accuracy. 

A measurement of \jd provides an alternative to the above approaches for obtaining \rrhosy . It requires a minimal amount of material (typically just a microbridge or nanobridge), does not require the complicated meander patterning needed for a kinetic-inductance measurement, 
 and is unaffected by a material's normal-state magnetism that affects inductive measurements of \rrhos as discussed above. This immunity to material magnetism was used to good advantage for directly obtaining \rrhos in the \ncco \ssc for the first time \cite{ncco}. Furthermore, unlike some of the methods for measuring \lm that do not provide an accurate absolute value but only provide the temperature variation $\lambda(T)/\lambda(0)$, \jd does provide the absolute value of \lm and \rrhos and can hence provide information on the total carrier concentration $n$. 

\subsection{Normal-State Resistivity}
Another valuable byproduct of measuring \jd is that it provides a direct measurement of the normal state resistivity \rrhon for temperatures below \tcy . One of the starting points in developing an understanding of any newly-discovered \ssc is to understand the underlying normal state: the type of carriers and their concentration, their band-related properties, and the relevant scattering mechanisms and rates. At low applied $B$ and \jjy , \rrhoy$(T)$ drops precipitously 
below \tcy, thereby obscuring how 
$\rho_n(T \ll T_c)$ would have behaved if the \scv had not set in. The most common method for measuring $\rho_n(T < T_c)$ utilizes high magnetic fields $B > B_{c2}$ to drive the system normal below \tcy; however, this measurement is subject to magnetoresistance (typically $R(B) \neq$ const) and may require prohibitively high magnetic fields ($B_{c2} >100$ T for some \sscy s). 

One alternative is to use the core of a magnetic flux vortex as a window to the normal state.
From Equation (\ref{fff}), a measurement of $\rho_f$ elucidates \rrhon \cite{metal}. However, this extraction of \rrhon requires interpretation and modeling, since Equation (\ref{fff}) usually holds only approximately except for very high driving forces, and the exact prefactor depends on the detailed regime of flux flow \cite{lo,blatter,hote}. 

Current-induced depairing provides an especially clean method for destroying \scv and 
accessing $\rho_n(T \ll T_0)$. Like the method of applying 
$B > B_{c2}$ to drive the system normal, applying $j > j_d$ is also free of 
interpretation and modeling, unlike flux-flow dissipation measurements. 
On the other hand, unlike the potential errors in the $B_{c2}$-based measurement due to normal-state magnetoresistance, the \jd method is immune to this issue because the normal-state electroresistance is quite negligible (i.e., $R(E) \simeq$ const) under the electric fields that arise at depairing conditions. 

Thus, besides the investigation of interesting phenomena and regimes in \scv related directly to current-induced depairing itself, the study of \jd provides information on the important parameters of the \scg state such as \rrhosy $(T)$ and \rrhony $(T)$.

\section{Relationship between the Depairing Current and Other Parameters} 
In a microscopic theory such as the
Bardeen--Cooper--Schrieffer (BCS) theory, experimental quantities are
calculated from microscopic parameters such as the strength of the
effective attractive interaction that leads to Cooper pair formation and
the density of states at the Fermi level. Often, these microscopic
parameters are not sufficiently well known. In the London and Ginzburg--Landau (GL) phenomenological theories, connections are made between the different observables from
constraints based on thermodynamic principles and
electrodynamical properties of the superconducting state, leading to an 
adequate estimation of the depairing current. These phenomenological formulations are described next \cite{tinkhamtext,pbreview}.

\subsection{London Formulation}
The London theory \cite{london,tinkhamtext} 
of superconductivity provides a description of the observed
electrodynamical properties by supplementing the basic Maxwell equations
by additional equations that constrain the possible behavior to reflect
the two hallmarks of the superconducting state: perfect conductivity and the
Meissner effect. Note that these properties hold only partially
when vortices are present. 

An ordinary metal (normal conductor) requires a driving electric field $E$
to maintain a constant current against resistive losses. 
In the simple Drude picture, this produces Ohm's
law behavior, $j=\sigma E$, with a conductivity given by 
$\sigma=ne^{2}\tau/m^{*}$. 
A superconductor can carry a resistanceless
current, and so, an electric field is not required for maintaining a persistent current.
Instead, $E$ in a perfectly-conducting state causes a ballistic
acceleration of charge so that: 
\be \label{london1}
E=\left( \frac{m^{*}}{n_{s}e^{2}}\right) \frac{\partial j}{\partial t} 
\ene
This is the first London equation, which 
reflects the dissipationless acceleration of the superfluid. 

The second property that needs to be accounted for is the expulsion of
magnetic flux by a superconductor. The magnetic field 
is exponentially screened from the interior following a spatial dependence: 
\be \label{field-decay}
\nabla^{2}B = B/\lambda_L^{2}
\ene
Together with the Maxwell equation $\nabla \times B = \mu_0 j$, 
this implies the following condition between $B$ and $j$:
\be \label{london2}
B= - \mu_0 \lambda_L^{2} 
(\nabla \times j)
\ene
This is the second London equation, which describes the property of a
superconductor to exclude magnetic flux from its interior. 
Taken together with the Maxwell equation $\nabla \times E = - \partial B/\partial t$, 
Equations (\ref {london1}) and (\ref{london2}) yield the expression for $\lambda_L$ of Equation (\ref{Ll}). 

Besides the London equations themselves, a third ingredient needed for the estimation of \jd in this framework is 
the thermodynamic critical field \hc and its relationship to the Helmholtz free energy density $f$. 
When flux is expelled, the free energy density is raised by the amount
$B^{2}/2\mu_0$. The critical flux expulsion energy 
(for the ideal case of a type-I superconductor with a
non-demagnetizing geometry and dimensions large compared to the
penetration depth) corresponds to the condition:
\be \label{fc}
f_{c} = f_{n} - f_{s} = \frac{B_{c}^{2}}{2\mu_0} 
\ene
where the L.H.S. of the equation represents the condensation energy density, which is the 
difference in free energy densities $f_{n} - f_{s}$ between the normal
and superconducting states. 
\jd represents the condition when the kinetic energy density equals the condensation
energy density: 
$\frac{1}{2} n_{s}m^{*}v_s^{2} 
=\frac{m^{*}j_{d}^{2}}{2 n_{s} e^{2}} = \frac{B_{c}^{2}}{2\mu_0}$, where $v_s$ is the superfluid 
speed. Substituting for $\lambda_L$ (Equation~(\ref{Ll})) gives the London estimate for the
depairing current density:
\be
\label{london-jd}
j_{d} \leq \frac{B_{c}}{\mu_0\lambda_L}
\ene
The inequality reflects the fact that $n_{s}$ does not remain constant, but diminishes 
as $j$ approaches \jdy . 

\subsection{Ginzburg--Landau Formulation} 
There are situations where a system's quantum wavefunction cannot be solved for
by usual means because the Hamiltonian is unknown or not easily approximated. 
The GL formulation \cite{GLpaper} is a clever construction that allows 
useful information and conclusions to be extracted 
in such a situation where one cannot solve the problem quantum mechanically.
For describing macroscopic properties, such
as \jd that we are about to calculate, the GL theory is in fact 
more amenable than the microscopic \linebreak theory \cite{tinkhamtext,bardeen}. 

The idea is to introduce a complex phenomenological
order parameter (pseudo wavefunction) $\psi=|\psi|e^{i \varphi}$
to represent the superconducting state. $|\psi (r)|^{2}$ is
assumed to represent the order parameter $\Delta$ introduced earlier, and to approximate the 
local density of paired superconducting charge carriers (Cooper pairs), 
which in turn is half the density of superconducting electrons $n_s$.

 The free energy density $f_{s}$ of the superconducting state is
then expressed as a reasonable function of $\psi (r)$ plus other energy terms.
A ``solution'' to $\psi (r)$ is now obtained by 
the minimization of free energy rather than through quantum mechanics.
The unknown parameters of the theory are then solved in terms of measurable
physical quantities, thereby providing constraints between the different
quantities of the superconducting state. 

Close to the phase boundary, \psay $^{2}$ is small, and so, $f_{s}$ can be
expanded keeping the lowest two orders of \psay $^{2}$. 
First, let us consider the simplest situation where there are no
currents, gradients in \psay , or magnetic fields present. Then, we have: 
\begin{equation} \label{simpleGL}
f_{s}= f_{n} + \alpha |\psi|^{2} + \frac{\beta}{2} |\psi|^{4}, \end{equation}
where $\alpha$ and $\beta$ are temperature-dependent coefficients whose
values are to be determined in terms of measurable parameters. 
The coefficients can be determined as follows. First of all, for the solution
of $|\psi|^{2}$ to be finite at the minimum free energy, $\beta$ must
be positive. Second, for the solution of $|\psi|^{2}$ to be non-zero,
$\alpha$ must be negative. Since $|\psi|^{2}$ vanishes above \tcy , 
$\alpha$ must change its sign upon crossing \tcy . 
The minimum in $f_{s}$ occurs at:
\be \label{GLsol1}
 |\psi|^{2}=-\alpha/\beta. 
\ene
Substituting this back in Equation~(\ref{simpleGL}) and using the definition of
\hc (Equation~(\ref{fc})), Equation~(\ref{GLsol1}) can be written as: 
\be \label{hc-alpha}
f_{c} = \frac{B_{c}^{2}}{2\mu_0} = \frac{\alpha^{2}}{2\beta} 
\ene
giving one of the connections between $\alpha$ and $\beta$ and 
a measurable quantity (\hcy ). 
A second connection can be obtained by noting
that $n_{s}$ in Equation~(\ref{Ll}) can be replaced by 2\psay $^{2}$, taking its
equilibrium value from Equation~(\ref{GLsol1}):
\be \label{lambdaGL}
\lambda^{2} = \frac{m^{*}}{2 \mu_0 |\psi|^{2} e^{2}} = 
\frac{-\beta}{\alpha}
\left(\frac{m^{*}}{2 \mu_0 e^{2}}\right)
\ene
Solving Equations~(\ref{hc-alpha}) and (\ref{lambdaGL}) simultaneously gives the 
GL coefficients:
\be \label{alpha-beta}
\alpha = -\frac{2 e^2 B_{c}^{2} \lambda^{2} }{m^{*}} 
\mbox{\hspace{2em} and 
\hspace{2em}} 
\beta=\frac{4 \mu_0 e^{4} B_{c}^{2}\lambda^{4} }{m^{*2}}
\ene
Note that $e$ and $m^*$ refer to single-carrier values and not pair values. 

To calculate \jdy , we include the effect of a current in Equation (\ref {simpleGL}) by adding
a kinetic energy term $\frac{1}{2} n_s m^{*} v_s^{2} = |\psi|^{2} m^{*} v_s^{2}$ to it: 
\begin{equation} \label{simpleGLfree}
f_{s} = f_{n} + 
\alpha |\psi|^{2} + \frac{\beta}{2} |\psi|^{4} + 
|\psi|^2 m^{*}v_{s}^{2}. \end{equation}
For zero \jj and $v_{s}$, we saw earlier (Equation~(\ref{GLsol1})) that 
the equilibrium value of $|\psi|^{2}$ that minimizes the free
energy is $|\psi_{j=0}|^{2}=-\alpha/\beta$. For a finite \jj and $v_{s}$,
minimization of Equation~(\ref{simpleGLfree}) gives the value of $|\psi|^{2}$ 
when it is suppressed by a current: 
\begin{equation}
|\psi_{j \neq 0}|^{2} 
=\frac{-\alpha}{\beta} \left( 1 - \frac{m^{*}v_{s}^{2}}{|\alpha|} \right) 
= |\psi_{j=0}|^{2} 
\left( 1 - \frac{m^{*}v_{s}^{2}}{|\alpha|} \right) \end{equation}
The corresponding supercurrent density 
is then: 
\begin{equation} \label{nonmonoj}
j= 2e |\psi_{j \neq 0}|^{2} 
v_{s}
= \frac{-2e \alpha}{\beta}
\left(1 - \frac{m^{*}v_{s}^{2}}{|\alpha|} \right) v_{s} 
\end{equation}
The maximum possible value of this expression can now be identified with
$j_{d}$: 
\begin{equation} 
\label{SBGL-jdt1}
j_{d}(T)= \frac{-4e \alpha}{3\beta}
\left(\frac{|\alpha|}{3m^{*}} \right)^{1/2} = 
\left(\frac{2}{3} \right)^{3/2} \frac{B_{c}(T)}{\mu_0 \lambda (T)}
\end{equation}
where the GL-theory parameters were replaced by their 
expressions in terms of the physical measurables \hc and $\lambda$ 
through Equation~(\ref{alpha-beta}). 
As anticipated at the end of Equation (\ref{london-jd}) for the London derivation for \jdy , that simpler 
estimate is indeed larger than this more rigorous GL derivation by the 
factor $(3/2)^{3/2}$=1.84.

The approximate temperature dependence of \jd can be obtained by inserting
the generic empirical temperature dependencies 
$B_{c}(T) \approx B_{c}(0)[1-(T/T_{c})^{2}]$ and 
$\lambda(T) \approx \lambda(0)/\sqrt{[1-(T/T_{c})^{4}]}$, giving:
\begin{equation} \label{jdtfull} 
 j_{d}(T) \approx j_{d}(0) [1-(T/T_{c})^{2}]^{\frac{3}{2}} 
[1+(T/T_{c})^{2}]^{\frac{1}{2}} 
\end{equation}
which close to \tc reduces to: 
\begin{equation} \label{jdtfull-dirty} 
 j_{d}(T) \approx \sqrt{2} j_{d}(0) [1-(T/T_{c})^{2}]^{\frac{3}{2}}. 
\end{equation}
where $j_{d}(0)$ is given by Equation (\ref{SBGL-jdt1}) by setting $T=0$
(for high scattering ``dirty'' \sscy s, the $\sqrt{2}$ prefactor can be smaller or absent
\cite{bardeen,romijn}). 

Since $B_{c}$ is not an easy quantity to measure directly, 
the relation: 
\be \label{hc-hc2-lambda}
B_{c}=\sqrt{\frac{\Phi_{0} B_{c2}}{4 \pi \lambda^{2}}}
\ene
can be used along with Equation (\ref{SBGL-jdt1}) to write the expression for \jdy $(0)$: 
\be \label{jdzeroGL}
j_{d}(0)=\sqrt{ \frac{2 \Phi_{0} B_{c2}(0) }{ 27 \pi \mu_0^2 \lambda^{4}(0) }}
\ene
that has the more easily measurable \hcuy . Since both \hcu and \jd can be obtained from transport measurements, this becomes a convenient way to obtain \lm and, hence, \rrhosy . 

\subsection{Microscopic Formulations and Generalizations} 
Various authors have calculated \jdt from a microscopic basis
\cite{bardeen,maki,ovchinnikov}. For
arbitrary temperatures and mean free paths, 
one must use the Gorkov equations as the starting point. Kupriyanov and
Lukichev \cite{KL} have derived \jdt from the Eilenberger equations,
which are a
simplified version of the Gorkov equations. This derivation is beyond
the scope of the present review, but a nice shortened version can be
found in \cite{romijn}. The microscopic calculation
confirms the overall temperature dependence predicted by GL, and
the two normalized curves differ only slightly from each other 
(e.g., see Figure~4 of \cite{romijn}). Thus, the GL theory can be applied 
over the entire temperature range down to $T \ll T_c$.
The previous equations relating \jd to \hc and \lm are expected to hold in the case
of multiple bands and other gap symmetries, as long as one uses 
the actual empirical temperature dependencies of \hc and \lmy , which 
account for modifications in these unconventional cases. This was experimentally demonstrated in the 
case of \mgb \cite{pbreview}, which was recognized as a multi-gap superconductor tuned by strain and doping 
in the early part of this century; in fact, \mgb showed superconductivity 
near a Lifshitz transition as in iron-based superconductors \cite{bauer, agrestini, kagan}. 

\section{Pulsed Measurement Technique}
Depairing current densities in \sscy s is extremely high: on the order of \jdy ($T$=0) = $10^{11}$--$10^{13}$ A/$m^2$. If the cross-section of the sample is even as narrow as just 1 mm$^2$, the current required would reach a value of $I=jA \sim 10^{6}$ A. Such a magnitude of current would be exceedingly difficult to produce and control. There are three steps to overcoming this dilemma: (1) Fabricate samples with very narrow cross-sectional areas. This can be achieved by growing nanowires and nanorods or by depositing very thin films and using lithography to pattern narrow bridges (alternatively, the films can be deposited onto nanowires or carbon nanotubes). (2) The next step is by pulsing the current at very low duty cycles so that large values of $I$ can be handled while reducing the time-averaged current and time-averaged power dissipation to manageable levels. (3) The last step is limiting the measurement of \jd to the regime close to \tcy. From Equation (\ref{jdtfull-dirty}), it would seem that \jd can be made arbitrarily small by making $T$ very close to \tcy; however, the $T -T_c$ distance needs to be large compared to the transition width for the measurement to be meaningful. Even for this near-\tc measurement of \jdy , the current usually will have to be pulsed to avoid significant sample heating. Furthermore, the near-\tc measurement will only measure \rrhos in that region, and its zero-$T$ value will have to be extrapolated using theory. While this is better than nothing, it will not shed light on any abnormal temperature dependence of \rrhos over the entire range, which could be of special interest if the superconductor has some exotic behavior. 

Thus, the experimental ingredients needed to conduct a \jd measurement are: a \scg sample with a very narrow cross-section; a means to control the temperature, i.e., a cryostat; and a method for sourcing pulsed signals (current or voltage) and detecting the consequent complementary signal (voltage or current). There are numerous methods for sample fabrication, which vary widely with the different \scg materials. Some deposition systems for preparing \scg films can be bought off the shelf. Cryostats also represent standard equipment that can be bought off the shelf. The principal distinguishing the experimental capabilities of our work center on the pulsed electrical measurements. Therefore, the rest of the experimental section will be devoted to describing this unique measurement setup.

\begin{figure}[ht]

\includegraphics[width=0.85\hsize]{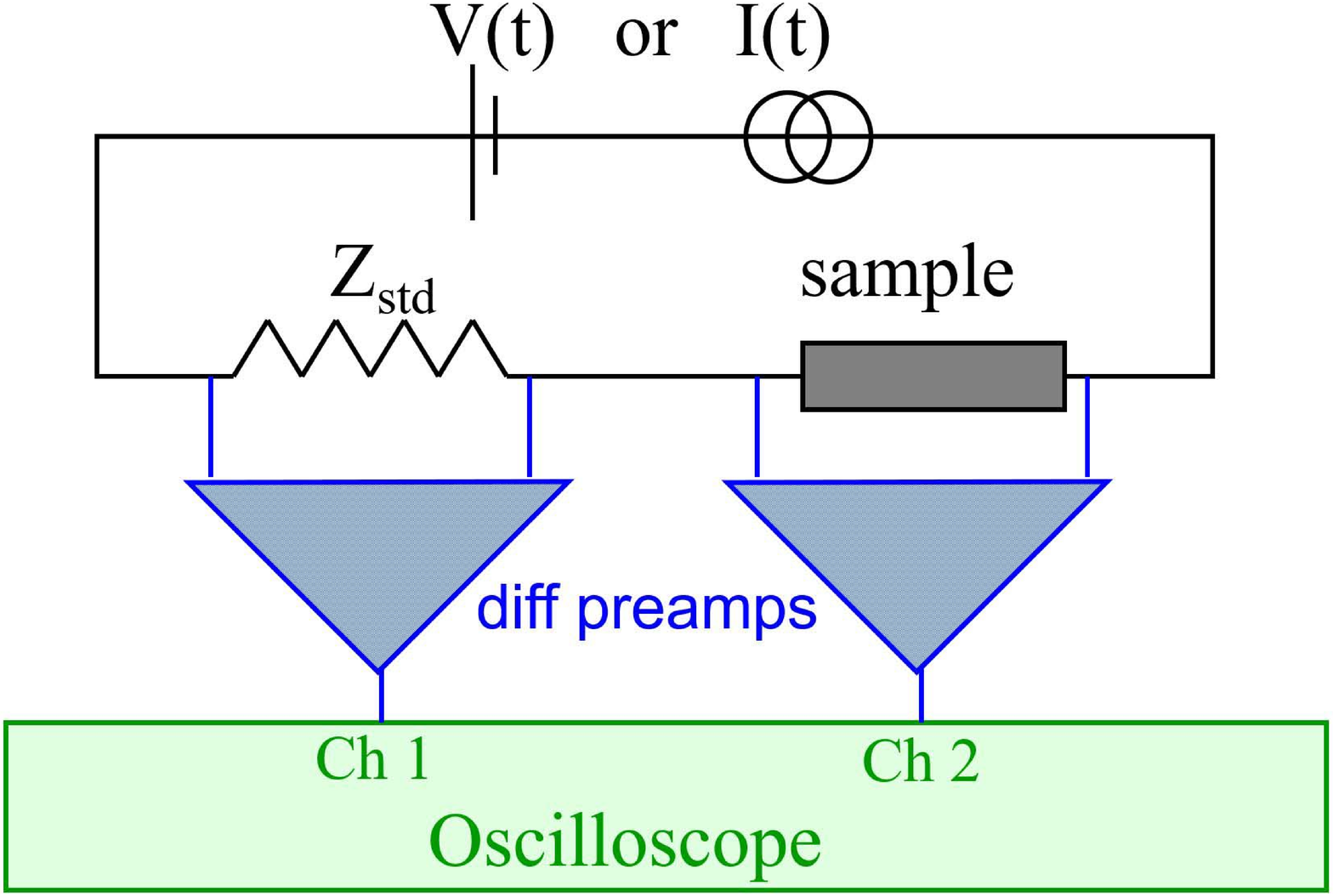} 
\caption{The overall configuration and functional schematic of the pulsed-signal measurement system. The differential preamplifiers (diff preamps) convert the time-varying potential differences across standard impedance ($Z_{std}$) and the sample, $V_{std}(t)$ and $V(t)$ respectively, into ground referenced single-ended signals that can be fed to the inputs of a digital storage oscilloscope.}
\label{pulsed-setup} 
\end{figure}

\figr{pulsed-setup} shows the overall configuration and functional schematic. The pulsed current/voltage source puts out a time-varying current and voltage. This signal flows through a standard impedance, usually a resistor \rstd (although an inductor is preferable in some situations) and the \scg sample of resistance $R$ that are in series. The initial signal can be taken directly from the output of a standard pulse generator (one of the models used was a Wavetek Model 801). These signal generators will typically have an output impedance of $Z_{out} = 50 \ \Omega$. If a lower $Z_{out}$ is desirable (to allow for constant voltage control), the signal generator's output can be passed through any standard buffer amplifier (e.g., a transistor-emitter-follower-based circuit, a power-operational-amplifier-based circuit, or an off-the-shelf audio amplifier). If a higher voltage than the signal generator's output is desirable, its output can be passed through any standard voltage amplifier (fast high quality audio amplifiers can serve this purpose as well). Combining a higher voltage signal with a large series resistor (which can be the \rstd itself or an additional series resistor) can provide a relatively constant current. In general, the measurement will be in current-controlled or voltage-controlled mode depending on whether the combination of the final 
$Z_{out}$ (after the amplifier if any) plus \rstd is greater than or less than $R$. If the current needs to be held constant to a high accuracy (for example, if a series of $R$ vs. $T$ resistive transition curves needs to be traced out at various constant currents, as will be seen later), then it is better to follow the pulse generator with a transconductance amplifier, which converts the generator's voltage pulse into a constant current pulse. The transconductance amplifier is able to hold the current constant by electronic circuity instead of needing an enormous series resistance. While conducting a pulsed current-voltage ($IV$) curve, which is usually done manually, it is preferable to have the voltage-controlled mode. The reason for this is that as the current and voltage are pushed higher, the sample's resistance will increase, and at some point, the sample will be driven to normal as \jd is exceeded. In constant-current mode, the power dissipation $P=I^2R$ rises as $R$ rises, causing an increase in heating and a further rise in $R$. This can lead to a run-away condition, which can destroy the sample. On the other hand, the constant-voltage mode is self-stabilizing since in this case, $P=V^2/R$ decreases as $R$ rises, thus reducing heating and controlling the situation.

Once the current pulse flows through the sample and \rstdy , the corresponding time-varying voltages, $V(t$) and $V_{std}(t)$, will be developed across them respectively. These must be observed and quantified using an oscilloscope. A digital storage oscilloscope (DSO) allows multiple pulses to be averaged. Since the signal is exactly repetitive, because the DSO is triggered off of the pulse generator's sync signal, the averaging effectively suppresses random uncorrelated noise. As long as the sample condition ($T$, $B$, etc.) is stable, a very high number of averages can be taken to improve the signal-to-noise ratio (SNR) vastly. Coaxial cables with 50-Ohm characteristic impedance are used between all connection points, including the wiring within the cryostat. Where possible, the originating and/or terminating points at the ends of the cables need to have matching 50-Ohm values to avoid reflections. Multiple ground connections to the circuit must be avoided to prevent ground loops. This means the two channels of the DSO cannot be simultaneously connected to both the sample and \rstdy; either a differential instrumentation preamplifier (Princeton Applied Research and Stanford Research Systems are two brands that make instrumentation amplifiers) must be used between the DSO channels and the sample and \rstdy, or only one of the two must be measured at a time. 

\begin{figure}[ht]
\includegraphics[width=0.85\hsize]{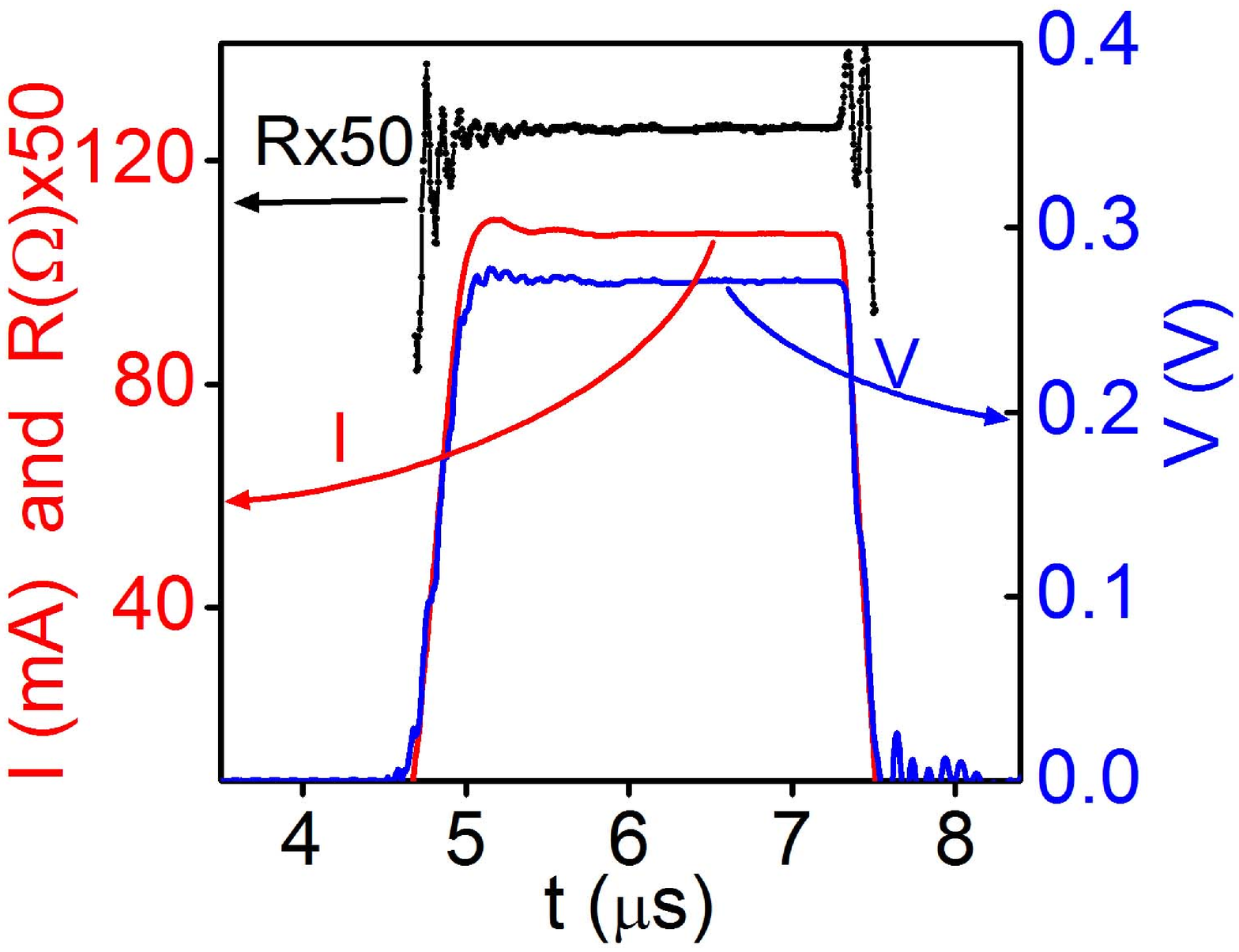} 
\caption{The measured oscilloscope traces of the sample voltage
$V(t)$, current $I(t)$=$V_{std}(t)/R_{std}$, and scaled calculated resistance $50R(t) = 50V(t)/I(t)$
for an MgB$_2$ bridge at 42 K (normal state just above \tcy). The resistance rises from 10\%--90\% of its final value in about 50 ns 
(adapted from reference \cite{pbreview}).}  
\label{pulses} 
\end{figure}

\figr{pulses} shows the pair of time-varying current 
$I(t)$=$V_{std}(t)/R_{std}$ and sample-voltage $V(t)$ 
signals that results. The topmost trace is the scaled calculated resistance 
$50R(t) = 50V(t)/I(t)$. Note that the pulses reach constant plateaus after their initial transients. $R$, $V$, and $I$ are defined by taking the plateau values of the individual quantities. The thermal rise in a sample because of Joule heating involves several processes: thermal diffusion occurs within the sample
essentially instantaneously; on the time scale of nanoseconds,
phonons transfer heat across the interface between the film and 
substrate; heat then diffuses within the substrate in a matter of 
microseconds and finally into the heat sink in milliseconds.
For those processes that have time scales comparable to or longer than the pulse duration, there will be a visible rise in $V(t)$, causing the pulse to be distorted. Thus, as long as the $V(t)$ pulse is flat, slow causes of heating that influence the $V(t)$ shape can be assumed to be negligible. The work in \cite{pair} discusses a method to evaluate a sample's thermal resistance for pulsed signals quantitatively. 

\begin{figure}[ht]
\includegraphics[width=0.85\hsize]{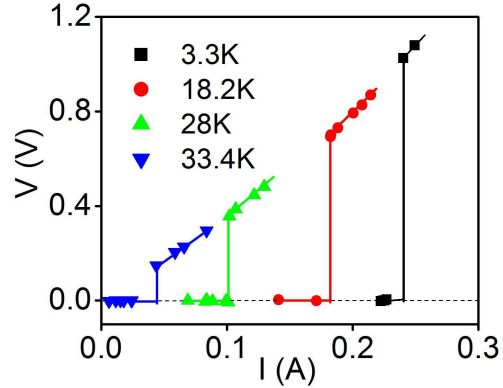} 
\caption{Current-voltage curves in a \mgb bridge at various fixed temperatures. As the temperature is increased, \jd is reduced, and hence, the ``jump'' occurs at a lower value of $I$. The sloped portions above each jump represent the normal-state resistance $R_n = \Delta V/\Delta I$ (adapted from from reference \cite{pbreview}).} 
\label{Mgb2-IVs} 
\end{figure}

\figr{Mgb2-IVs} shows an example \cite{pbreview} of a set of IV curves at various fixed temperatures (in zero magnetic field), where each data point represents a pulsed measurement (plateau values) as described above. As the temperature is increased, \jd is reduced, and hence, the ``jump'' occurs at a lower value of $I$. Notice that the resistance (the V/I slope) jumps from zero (dissipationless \scg state) to a constant finite value (normal-state) as the current crosses its depairing value. This is one direct way of obtaining \rrhon below \tcy. In this particular material, high impurity scattering dominates over electron-phonon scattering at all temperatures, leading to a relatively flat $\rho(T)$. A more interesting application of this technique for elucidating a variable $\rho(T)$ is described in a later section. 

\begin{figure}[ht]
\includegraphics[width=0.85\hsize]{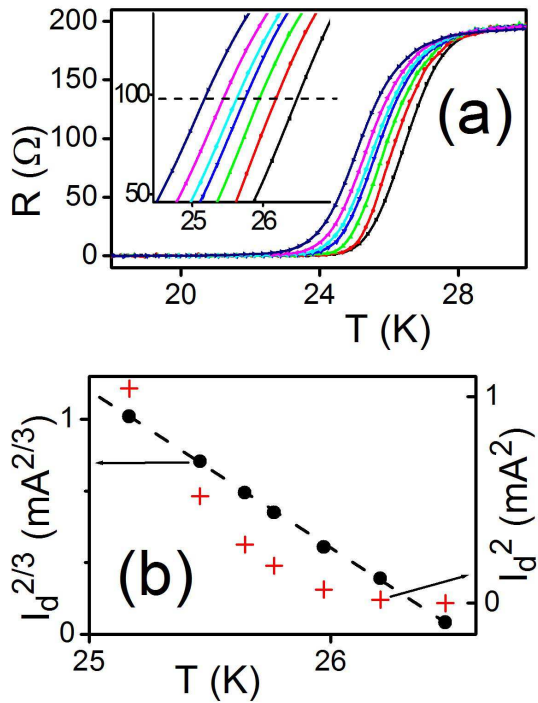} 
\caption{(\textbf{a}) Resistive transitions in a Sr$_{1-x}$La$_x$CuO$_2$ \scg film bridge in zero magnetic field at various transport current values (right to left): 12.9, 132, 258, 426, 533, 721, and 1020 $\mu$A. The lowest current is
continuous DC; the remaining currents are pulsed. The inset shows a magnified view of the midpoint region. (\textbf{b}) Left axis (circle symbols): two-thirds power of the
depairing current versus the midpoint transition temperature corresponding to the depairing law. Right axis (plus symbols): square of the depairing current versus the midpoint transition temperature corresponding to Joule heating 
(adapted from reference \cite{slcojd}).}
\label{SLCOjd} 
\end{figure}

\figr{SLCOjd}a represents a set of pulsed constant-current $R(T)$ curves in zero magnetic field \cite{slcojd}. As the current is increased, the transition is progressively pushed down in temperature. \figr{SLCOjd}b plots these midpoint transition temperatures ($T_{c2}$) versus $I$, and they are seen to follow a $I^{2/3}$ law as per Equation (\ref{jdtfull-dirty}). From this measured slope and Equation (\ref{jdtfull-dirty}), one can estimate $j_d(0)$ without requiring the application of this enormous value of current. This is especially useful for systems (e.g., cuprate high-temperature \sscy s) that have a very high $j_d(0)$ value. 

\begin{figure}[ht]
\includegraphics[width=0.85\hsize]{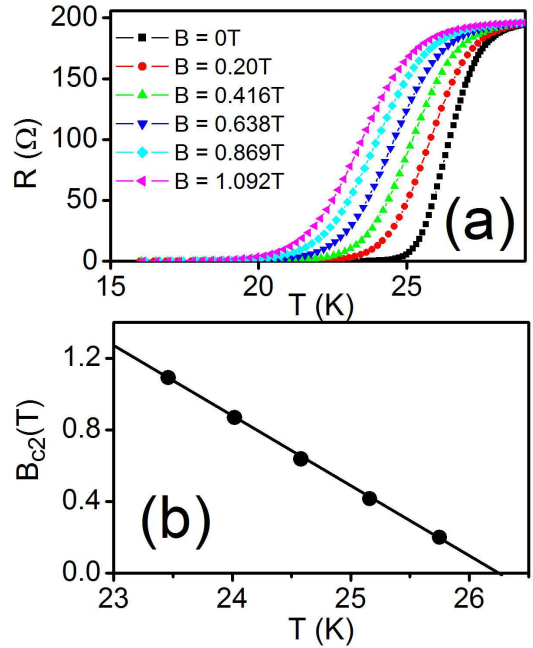} 
\caption{ (\textbf{a}) Resistive transitions of a Sr$_{1-x}$La$_x$CuO$_2$ \scg film bridge at a constant current of I = 13 $\mu$A in various perpendicular magnetic field values as indicated in the key. 
(\textbf{b}) Upper critical magnetic field versus the midpoint transition temperature, extracted from the curves in ({a}) 
(adapted from reference \cite{slcojd}).}
\label{SLCO-Bc2}
\end{figure}

\figr{SLCO-Bc2} shows the companion very low DC-current $R(T)$ curves at various constant $B$ values. Here, the shift occurs because of the $B_{c2}(T)$ boundary; the current level is small enough for its depairing to be negligible. Unlike $j_{d}(T)$, $B_{c2}(T)$ has a linear dependence near \tc , and that slope can be related to $B_{c2}(0)$ through the WHH (Werthamer, Helfand, and Hohenberg) theory \cite{whh} and its variations \cite{gurevich-Bc2} by relationships such as $B_{c2}(0) \simeq 0.7 dB_{c2}/dT$. 

The measurements represented by \figr{SLCOjd} and \figr{SLCO-Bc2} together with Equation (\ref{jdzeroGL}) are the key to obtaining \rrhos through relatively straightforward transport measurements. We now look at one recent example of a current-induced depairing study of an exotic superconducting system.

\section{Investigations in a Topological Insulator/Chalcogenide Interfacial Superconductor }
\subsection{Background}
The interface between the Bi$_2$Te$_3$ topological insulator and the FeTe chalcogenide provides a fascinating 2D \scg system, in which neither Bi$_2$Te$_3$, nor FeTe are \scg by themselves \cite{he}. 
While the exact origin of the \scv is not known, it has been suggested that 
the robust topological surfaces states 
may be doping the FeTe and suppressing the antiferromagnetism in a thin region close to the interface, thus inducing the observed 2D \scvy . 
These surface states represent a conducting 
system with very high normal conductivity because of protection against time-reversal-invariant scattering mechanisms. Therefore, it is of great interest to understand the nature and origin of the charge carriers that underlie this interfacial \scvy , and in particular, to see if the topologically-protected surface states might be a source of the normal carriers. The relevance of this question is broader than the specific system studied here, since it has been recently proposed that interfacial superconductivity may even play a role in cuprates: for example, in the interface located between charge density wave nanoscale puddles \cite{campi2015} and between oxygen-rich grains where the interface is made of a filamentary network with 
hyperbolic geometry \cite{jarlborg2019,campi2016}. We describe below how the current-induced depairing approach was used to answer these questions to elucidate the nature of the normal state in the \bite system. 

\subsection{Samples and Experimental Information}
The \bite samples consisted of a ZnSe buffer layer (50 nm) deposited on
a GaAs (001) semi-insulating substrate, followed by a deposition of 
220 nm thick FeTe, which was then capped with a 20 nm-thick \bt layer. 
Upper-critical-field measurements \cite{he} and vortex-explosion \linebreak measurements \cite{FTBTVortexExplosion} showed that the \scv occurred within an interfacial layer of thickness $d = 7$ 
nm, which was much thinner than both the FeTe and \bt layers.
Projection photolithography followed by argon-ion milling was used to pattern narrow microbridges optimized for the high current-density pulsed four-probe measurements. Two bridges were studied: Sample A with lateral dimensions of 
width $w=11.5 \ \mu$m and length $l=285 \mu$m and Sample B with $w=12 \ \mu$m and $l=285 \mu$m. 
The onset \tc (defined as the intersection of the extrapolation of the normal-state portion and the extrapolation of the steep transition portion of the $R(T)$ curve) for both bridges, was 11.7 K. Details about sample preparation are provided in~\cite{he}.
All measurements were made in zero applied magnetic field.
While the very low reference curves at $I \leq 60 \ \mu$A were measured using continuous DC signals, the main electrical transport measurements were made with pulsed signals. 
Contact resistances ($< 1 \ \Omega$) were much lower than the normal resistance
$R_{n}$ of the bridge, and heat generated at contacts did not reach 
the bridge within the time duration $t$ of each pulse, since the thermal diffusion 
distance ($\sqrt{Dt} \sim 10\mu$m) was much shorter than the contact-to-bridge distance ($>1$ mm); here, $D$ is the diffusion constant. 

\subsection{Results and Discussion} 
\begin{figure}[ht]
\includegraphics[width=0.85\hsize]{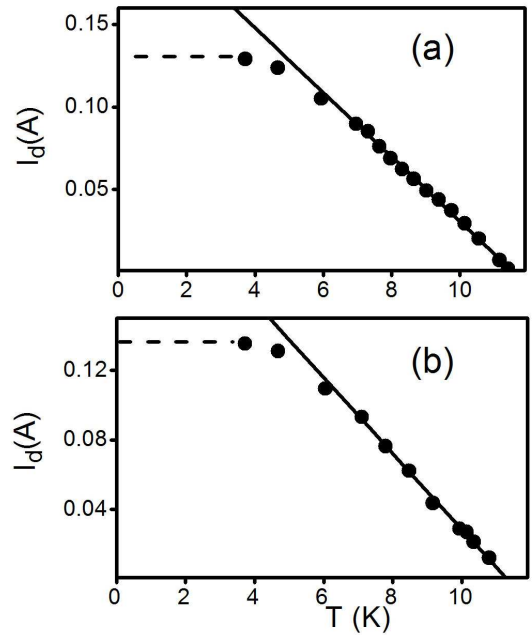} 
\caption{Raw depairing current versus temperature for \bite bridges: (\textbf{a}) Sample A and (\textbf{b}) Sample B.
The dashed horizontal lines provide the values $I_d(T\rightarrow 0) \geq 0.131$ A and 0.136 A for Samples A and B, respectively 
(adapted from reference \cite{jdbite}).}
\label{jdbite_Id} 
\end{figure}

The normal-state resistivity $\rho_n (T)$ and depairing current density $j_d(T)$ in the \bite samples were extracted 
over the entire temperature range \cite{jdbite}, by driving the system normal with high pulsed currents using the methods described earlier and illustrated in \figr{Mgb2-IVs} and \figr{SLCOjd}. \figr{jdbite_Id} shows the raw depairing current results. The dashed horizontal lines in Panels (a) and (b) provide the values $I_d(T\rightarrow 0) \geq 0.131$ A and 0.136 A for Samples A and B, respectively.

\begin{figure}[ht]
\includegraphics[width=0.85\hsize]{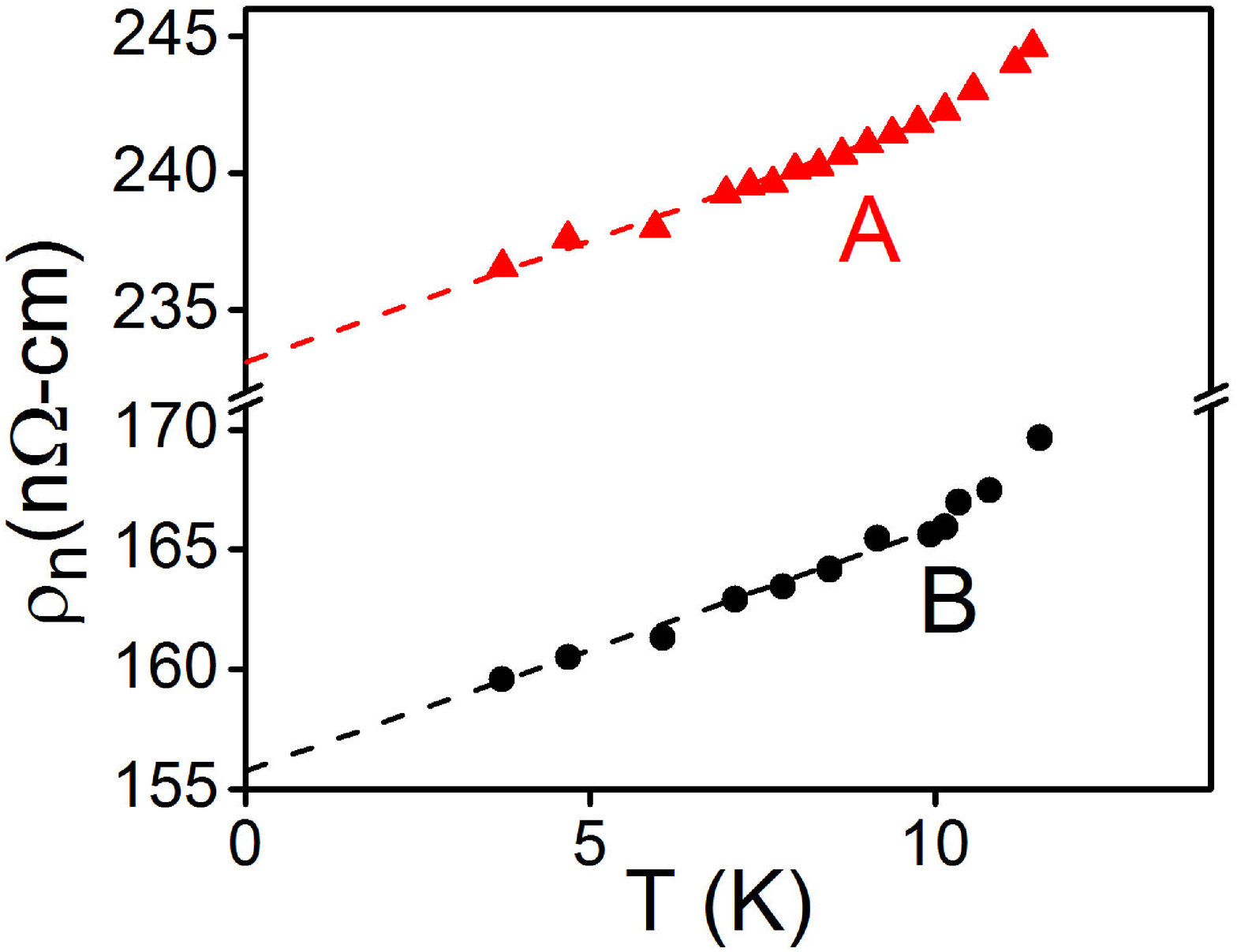} 
\caption{ 
Intrinsic normal-state resistivity of \bite interfacial \scg bridges, 
Samples A and B 
(adapted from reference \cite{jdbite}).}
\label{Rn-vsT} 
\end{figure}

In order to obtain more accurate intrinsic \jd and $\rho_n$ 
of the 7 nm-thick \sg interfacial layer itself, we needed to subtract the small parallel current through the normally conductive underlying FeTe layer. For this purpose, a separate measurement of pure FeTe deposited on ZnSe/GaAs, without the \bt top layer, 
was conducted \cite{jdbite}. With this subtraction, 
the previous raw $I_d(T\rightarrow 0)$ values gave a 
corrected \jdy $(T=0)$ of $1.5 \times 10^8$ A/\cms for both samples (which is a typical value: \jd ranges $10^7$--$10^9$ A/\cms for most superconductors), and the correction gave 
the intrinsic $\rho_n(T)$ for the two samples, as shown in \figr{Rn-vsT}. This absolute value of $\rho_n(T$$\rightarrow$$0)$$\sim$200 n
$\Omega$ cm represents an extraordinarily conductive 
normal state for a \sg system, as most superconductors are poor conductors in the normal state.
This information will be analyzed below within the framework of an anisotropic Ginzburg--Landau (GL) approach \cite{jdbite}, 
to obtain information on the superfluid density, carrier concentration, and 
scattering rate, as well as their implications for the nature of the normal-state.

From the previously-published measurements of He {et al.} \cite{he}, we have the following orientation-dependent 
values of \hcuy : perpendicular-to-interface $B_{c2}^{\perp}(0) \approx
17$ T and parallel-to-interface $B_{c2}^{\parallel}(0) \approx 40$ T. The corresponding coherence lengths from Equation (\ref{bc2-xi}) are: in-plane $\xi_{\parallel}(0) \approx 4.4$ nm and perpendicular $\xi_{\perp}(0) \approx 1.9$ nm. 
Using Equation (\ref{jdzeroGL}) together with this $B_{c2}^{\perp}(0)$ and our measured in-plane $j_d^{\parallel}(0)$ gave 
$\lambda_{\parallel}(0) = 124$ nm and a corresponding $\rho_s(0) = 1/\lambda^2_{\parallel}(0)$.
From $\rho_s(0) = \mu_0 n_s(0) e^{2}/m^* \approx \mu_0 n e^2/m$ 
applicable in the clean limit at $T=0$, we get $n \approx 1.8 \times 10^{21}$ per \cmcy , approximating $m^* \approx m$. This effective single-band value of $n$ evaluated above 
is similar to $n$ characteristic of high temperature superconductors and about two orders of magnitude lower than $n$ in highly-conductive metals such as copper. 

The low value of $n$ together with the very high normal conductivity 
implies a rather long scattering time $\tau$ and mean-free-path $l$. 
The Fermi wave number for this $n$ computes to 
$k_F(3D) = m^*v_F/\hbar = (3\pi^2 n)^{1/3} = 
3.8 \times 10^9$ m$^{-1}$ and $k_F(2D) = (2\pi n d)^{1/2} = 
9.0 \times 10^9$ m$^{-1}$ in three and two dimensions, respectively. In both cases, the Fermi wavelength $\lambda_F=
2\pi/k_F \ll d$, validating the continuum approximation for states along the perpendicular direction and 
 justifying the anisotropic 3D treatment of the normal state. Then, from the Drude relationship 
$\rho \approx m/n e^2\tau$, we get $\tau \approx 10$ ps, 
which agrees well with the scattering rates ($\sim \hbar/0.05$ meV = 13 ps) measured by Pan et al. \cite{pan} 
using spin- and angle-resolved photoemission spectroscopy. 
Combining this value of $\tau$ with the Fermi velocity $v_F = \hbar k_F/m 
\approx 440$ km/s, we get $l= v_F \tau = 4.2 \ \mu$m. 
The very long $l$, which well exceeds the \scg layer thickness $d$, 
indicates that scattering from the faces that bound the \sg layer was of a specular nature. This surprising dramatically low scattering indeed supports the possible role of the topological surface states in the formation of the normal state that underlies this exotic interfacial superconducting system.

\section{Concluding Remarks}
Fast pulsed signals of short duration and low duty cycle make it possible to study transport behavior in \sscy s at extreme current densities, power densities, and electric fields. In this article, we focused on the use of this technique in the measurement of one of the fundamental critical parameters of the \scg state, the depairing current \jdy . It was shown how through \jd , one can obtain information on various other key parameters of the \scg state, in particular the penetration depth and consequent superfluid density, which cast light on the normal state. As an example and illustration of this procedure, we described a recent study of the \scg system formed at the interface between a topological insulator and a chalcogenide. We hope that the information provided here will encourage other groups to utilize this approach. 


\section{Acknowledgments}
The following are acknowledged for useful discussions and other assistance: Charles L. Dean, Manlai Liang, Gabriel F. Saracila, 
James M. Knight, Luc Fruchter, Ziang Z. Li, Qing Lin He, Hongchao Liu, Jiannong Wang, Rolf Lortz, Iam Keong Sou, Alex Gurevich, Richard A. Webb, Ken Stephenson, and David K. Christen.
This work was supported by the U.S. Department of Energy through Grant Number DE-FG02-99ER45763.


\end{document}